\documentstyle[12pt,epsf]{article}
\begin{document}

\hfill \vbox{\hbox{UCLA/99/TEP/25}
             \hbox{INLO-PUB-11/99}
             \hbox{hep-lat/9905029}} 
\begin{center}{\Large\bf On P-vortices and the Gribov problem
}\\[2cm] 
{\bf Tam\'as G. Kov\'acs}\footnote{Research supported by 
FOM.} \\
{\em Instituut-Lorentz for Theoretical Physics, 
P.O.Box 9506, 2300 RA, Leiden, The Netherlands}\\
{\sf e-mail: kovacs@lorentz.leidenuniv.nl}\\[5mm] 

and\\[5mm]

{\bf E. T. Tomboulis}\footnote{Research supported by 
NSF grant NSF-PHY 9531023}\\
{\em Department of Physics, UCLA, Los Angeles, 
CA 90095-1547}\\
{\sf e-mail: tombouli@physics.ucla.edu}
\end{center}
\vspace{1cm}

\begin{center}{\Large\bf Abstract}
\end{center}

We study the possible connection between centre vortices
and P-vortices in SU(2) gauge theory. After briefly recalling  
some essential properties of centre vortices we point out that
there is no known a priori connection between the gauge dependent
P-vortices and the gauge invariant centre vortices. We then 
show by Monte Carlo simulations that the `centre projected
physics' strongly depends on the gauge copy from which 
the maximal centre gauge fixing is started. This reveals 
the presence of Gribov problems, and casts some 
doubts on the physical meaning of P-vortices, and should be further 
investigated.

\vfill
\pagebreak  

Recently there has been considerable interest in the role
that vortices might play in confinement in non-Abelian 
gauge theories \cite{lat98}. 
One approach towards isolating such configurations in the 
vacuum which has been pursued by several authors 
\cite{GS-PRD} - \cite{forc}, \cite{vorgeom} is that of 
the so called centre projection and projection vortices. 
Centre projection involves a gauge fixing to the so called
maximal centre gauge (MCG) that puts the SU(N) gauge group valued
link variables as close as possible to elements of the
centre, Z(N). After gauge fixing the link variables
are `projected' onto the centre, i.e. they are replaced
by the closest centre element. This procedure is in complete 
analogy with Abelian projection in the maximal Abelian gauge.
Most of the results obtained with MCG so far are for the case 
of $N=2$ and we shall also mostly concentrate on this case.

The excitations of the resulting Z(2) gauge 
theory after the projection are very simple objects:
co-closed (their co-boundary is empty) sets of plaquettes 
each carrying -1 flux. In analogy with vortices in the full SU(2) theory, 
these are referred to as projection vortices or P-vortices. A detailed
study of full SU(2) Wilson loops with an even and odd number
of P-vortices linking with them showed that the sign of large
Wilson loops is strongly correlated with the parity of the number
of P-vortices linking with the loop \cite{GS-PRD}. 
Although no exact connection is known between P-vortices 
and vortices in the full SU(2)
system, this property led to the tentative
identification of P-vortices with the `cores' of full SU(2) vortices.
The centre projected configurations were 
found in \cite{GS-PRD} to carry the full SU(2) string tension, and,   
furthermore, the density of P-vortices 
was claimed to be a scaling physical 
quantity \cite{tubingen-scale,GS-PRD}. Further work in 
MCG and centre projection led to suggestions that P-vortices 
account for the finite temperature properies
of the SU(2) theory \cite{tubingen-T}, and that the   
elimination of P-vortices results in the loss of 
confinement and chiral symmetry breaking \cite{forc}.  

In this letter we would like to point out some potential 
problems with this picture and draw attention 
to some questions that in our view would
have to be clarified. We first point out that, 
although P-vortices are known in some cases to be correlated 
with centre fluctuations of Wilson loops,   
there is in fact no established connection between topological 
centre vortices in $SU(N)$ gauge theory and P-vortices. 
We then proceed to demonstrate some potential problems with
the gauge fixing and projection procedure that are due to 
the Gribov ambiguity. By presenting results of a Monte Carlo
computation, we show that if the centre gauge fixing is started
from already Lorentz gauge fixed configurations then on the
average the MCG fixing arrives to a higher local maximum 
of the gauge fixing functional than when started from a random
gauge; but at the same time the
projected configurations have only a very small fraction
of the full SU(2) string tension. Moreover, the density
of P-vortices which is claimed to be a scaling physical 
quantity, is significantly smaller than without an initial
Lorentz gauge fixing. This, when combined with the previously known 
fact that centre projection results are unstable under local 
smoothing \cite{GS-PRD}, casts doubts on the physical
relevance of P-vortices and calls for further investigations 
on whether P-vortices can be defined in an unambiguous manner,
so that the projection-physics be independent of the details
of the gauge fixing procedure.

{\it Centre vortices versus P-vortices} A vortex is a configuration 
of the gauge potentials ${\bf A}_\mu$ 
topologically characterised by nontrivial elements of 
$\pi_1(SU(N)/Z(N)) = Z(N)$. This means that over a sufficiently 
large loop the configuration is characterised by a singular gauge 
transformation which cannot be consistently defined throughout the 
space encircled by the loop without encountering a topological 
obstruction; the multivaluedness ambiguity is by elements of $Z(N)$. 
Note that the topological $Z(N)$ flux is conserved only 
mod $N$, so the number of vortices linked with a given loop is defined  
only mod $N$. 
On the lattice only configurations with one-plaquette action near its 
maximum contribute significantly at large $\beta$. In fact such 
local smoothness is essential in order to be able to talk of 
vortices relevant to the continuum limit in the first place; 
according to rigorous theorems, only for lattice 
configurations with sufficiently small plaquette function variations 
from the maximum is it possible to unambiguously define a continuum 
interpolation assignable to a topological sector. This implies that only 
locally smooth very extended vortices (`thick' vortices)  
are of interest. They incorporate potential long distance disordering 
together with UV asymptotic freedom. It was indeed found 
\cite{KT} that the induced centre fluctuations for 
large Wilson loops in $SU(2)$ and $SU(3)$ 
carry the full string tension, and in a manner which is perfectly 
stable under repeated smoothings removing short distance fluctuations. 

Now given such an $SU(2)$ vortex configuration, how can a P-vortex 
be `associated' with it? 
By considering regular gauge transforms 
of it, the nontrivial $Z(2)$ element of the 
singular gauge transformation asymptotically characterising the 
vortex may be made to be distributed over many or few links 
of an encircling loop. Assume that by a gauge transformation it is 
concentrated on just one. One may then try to 
extend this gauge transformation in directions 
perpendicular to the loop so that a jump by 
$-1$ occurs when crossing this `wall'. In essence one is trying 
to compress the thick vortex into a thin one by a gauge 
transformation. In the presence of the lattice cutoff, in 
the corresponding Z(2) projected configuration this may appear 
as a $Z(2)$ vortex (P-vortex) linking with the loop. However, 
it is clear that this procedure is ambiguous. 
A slightly different gauge transformation 
may distribute the $Z(2)$ jump over two or three links, in such 
a way that the $Z(2)$ projected configuration has one, three, or much worse, 
zero or two P-vortices linking with the loop. There is no guarantee that 
a gauge condition such as the MCG will always pick out one 
but not the other from two such configurations. Or that starting 
from gauges where the vorticity along a large loop is 
spread out as smoothly as possible, 
Gribov problems will not arise in trying to go over to  
centre gauges. At any rate, we are not aware of any well-defined 
procedure of associating the gauge-dependent concept of a P-vortex 
with the notion of an extended $\pi_1(SO(3))$ vortex in the gauge 
field. 

All this raises the possibility that 
variations in the centre gauge fixing procedure and Gribov 
ambiguities may lead to widely 
different number of P-vortices. Worse, there may be no stable mod $2$
correspondence between the number of P-vortices and the number of 
$SU(2)$ vortices. The string tension from P-vortices may then 
change substantially under similar gauge fixing schemes. 
We next proceed to investigate some of these possible problems.

{\it The Gribov problem} The most essential ingredient 
in the definition of P-vortices
is the gauge fixing to the maximal centre gauge (MCG) by maximizing
the functional
\begin{equation}
 f[U] = \sum_l (\mbox{tr} U_l)^2
\end{equation}
where the summation is over all the links of the lattice. After
gauge fixing, a ``projection'' is performed which amounts to the
replacement of all the link variables with the closest centre 
element. In the case of SU(2) this is simply a $U_l \rightarrow 
\mbox{sign tr}U_l$ replacement which results in a Z(2) gauge
configuration. The procedure is completely analogous to 
Abelian projection in the maximal Abelian gauge.

Unfortunately, in practice the gauge fixing is ambiguous
because the gauge fixing functional $f$ has several local
maxima over the whole gauge orbit. It is practically
impossible to select the absolute maximum of these local 
maxima, instead, a local maximum is achieved only. Gauge
non-invariant quantities can depend on which local maximum
is selected. A typical way of dealing with this problem is
to perform the gauge fixing on several randomly selected 
gauge copies of the same configuration and using only the 
one for which the highest maximum of $f$ is obtained 
\cite{GS-PRD}. In the present paper we would like to study how sensitive
the ``centre projected physics'' is to the Gribov ambiguity.  

Physically, the MCG tries to compress most of the fluctuations
into objects on the one lattice spacing scale, i.e.\ P-vortices.
On the other hand Lorentz gauge achieves exactly the opposite.
By putting all the links as close to the identity as possible,
it tries to spread the fluctuations evenly. The question we
would like to ask is whether there is a significant difference
in the performance of the MCG fixing depending on what type
of gauge copy we start it from. Therefore we made two gauge 
copies of the same set of configurations, one random copy
and another fixed to the Lorentz gauge. We then compared the
projection physics of these two equivalent ensembles.

We used a set of 100 $12^4$ SU(2) gauge configurations generated
with the Wilson action at $\beta=2.4$. The random gauge copies
were simply the ones resulting from the Monte Carlo and the
other ensemble was prepared by fixing the same set of 
configurations to the Lorentz gauge by maximising $\sum_l 
\mbox{tr}U_l$. We then performed MCG fixing and projection 
on both ensembles using the over-relaxation
algorithm described in \cite{GS-PRD}. The iteration was stopped 
when the change in gauge fixing action per degree of freedom 
was smaller than $1.0\times 10^{-9}$. 

\begin{figure}[!htb]
\begin{center}
\vskip 10mm
\leavevmode
\epsfxsize=120mm
\epsfbox{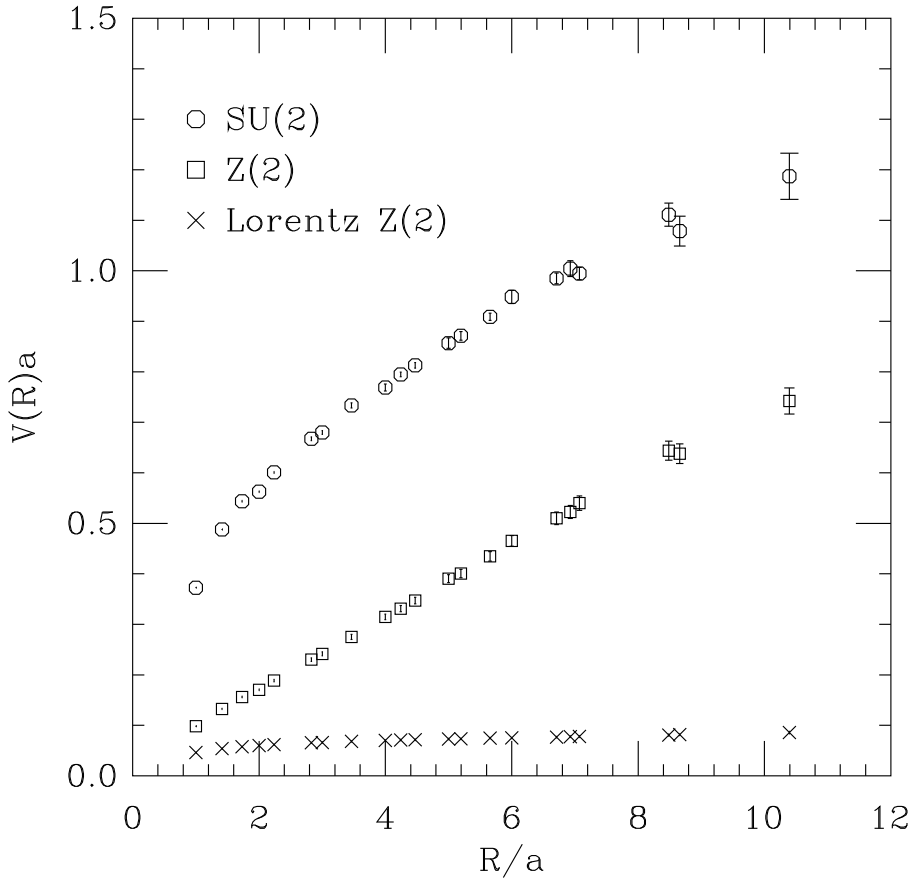}
\vskip 10mm
\end{center}
\caption{The heavy quark potential measured on a set of
100 $12^4$ Wilson $\beta=2.4$ gauge configurations. SU(2)
is the full SU(2) potential, Z(2) is measured on the 
centre projected configurations, and ``Lorentz Z(2)''
is measured on the first Lorentz gauge fixed then 
MCG fixed and then centre projected ensemble.}
\label{fig:potz2}
\end{figure}

In Figure \ref{fig:potz2} we compare the heavy quark
potential obtained from the two centre projected ensembles 
as well as the SU(2) configurations. As expected from the
work of Del Debbio et al.\ \cite{GS-PRD}, the projected
ensemble for which the gauge fixing was started from a 
random gauge, reproduces the full SU(2) string tension 
but without the coulomb term in the potential. On the
other hand, if the MCG fixing is started from already
Lorentz gauge fixed configurations, the resulting 
``projection physics'' is entirely different. The string 
tension is compatible with zero, there is no confinement.
We stress that the two projected ensembles producing 
so vastly different projection physics, should 
in principle be exactly equivalent, the two SU(2) ensembles
being gauge copies of one another.

It is also instructive to compare the average value of the
maximum of the gauge fixing function obtained for the two
ensembles. In the random gauge started ensemble it is 
3.0914(10), while in the Lorentz gauge started ensemble
it is 3.1012(8). In fact, the MCG fixing arrives at a 
higher local maximum if started from Lorentz gauge
and not a random gauge. 

It has been observed that the area-density of P-vortices
is a properly scaling physical quantity \cite{GS-PRD}.
This quantity is essentially the volume density of 
-1 plaquettes in the projected configurations, i.e., in
lattice units
\begin{equation}
  p = \frac{N_{vor}}{6 \times \mbox{Volume}},
\end{equation}
where $N_{vor}$ is the number of -1 plaquettes. In a recent
work the geometrical structure of these vortex surfaces on the
dual lattice has been studied in great detail \cite{vorgeom}.
We also measured the P-vortex density in both ensembles.
In the random gauge started ensemble we obtained $p=0.0552(5)$,
consistently with Ref.\ \cite{GS-PRD}. The Lorentz gauge
started ensemble produced a significantly smaller result,
$p=0.0338(2)$. Considering the complete absence of a string
tension in the latter ensemble, a drop of only about 40\% in
the vortex density indicates that the structure of these
remaining P-vortices must be vastly different from the ones
occurring in the random gauge started ensemble. They must be
small localized vortices, not contributing to the string tension.

\begin{figure}[!htb]
\begin{center}
\vskip 10mm
\leavevmode
\epsfxsize=120mm
\epsfbox{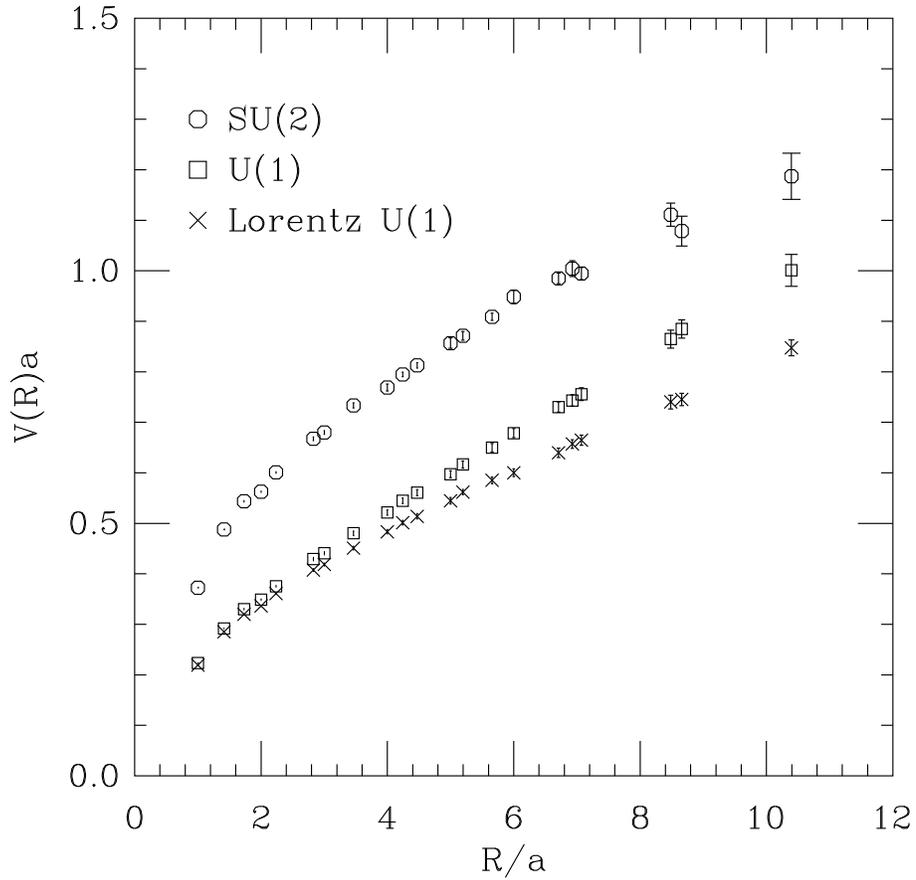}
\vskip 10mm
\end{center}
\caption{The heavy quark potential measured on a set of
100 $12^4$ Wilson $\beta=2.4$ gauge configurations. SU(2)
is the full SU(2) potential, U(1) is measured on the 
Abelian projected configurations, and ``Lorentz U(1)''
is measured on the first Lorentz gauge fixed then 
maximal Abelian gauge fixed and then Abelian projected ensemble.}
\label{fig:potU(1)}
\end{figure}

For comparison we repeated the above experiment using 
the same set of SU(2) gauge configurations but now instead
of centre projecting in the MCG, we used 
Abelian projection in the maximal Abelian gauge. In this 
case the Gribov problem is also potentially present. 
The maximum of the gauge fixing action however 
shows only a slight difference if any between the two ensembles.
We obtained 1.4620(4) for the ensemble started from a random
gauge and 1.4632(4) for the Lorentz gauge started ensemble.
Similarly, the heavy quark potentials, shown in Fig.\ 
\ref{fig:potU(1)} are also not as vastly different as in the 
case of the MCG. In fact, more extensive computations would be needed 
here in order to draw a definitive conclusion about the 
possible difference in the two string tensions.  
This indicates that the Gribov ambiguity probably does not 
influence the physical observables so severely in the
MAG case as it does in the case of the MCG.  

Some time ago it was noticed that a slight local smoothing of
the SU(2) configurations decreases the centre projected
string tension considerably \cite{GS-PRD}. 
This shows that SU(2) configurations
with similar long distance properties but with different
short distance fluctuations, can produce different
projection physics. Somehow the centre projection entangles
excitations living on different length scales and it is not
clear whether the centre projected configurations really reflect
the long distance SU(2) physics. 
Our present result is even more alarming. It shows that {\em
exactly gauge equivalent} SU(2) configurations, 
when fixed to the MCG and centre projected, can produce Z(2) 
configurations with vastly different physical properties. 
This casts doubts on the physical meaning of P-vortices
and certainly calls for a more detailed study of the Gribov
problem in the context of centre projection.

\section*{Acknowledgements}

TGK thanks Pierre van Baal and Margarita Garc\'{\i}a P\'erez for
discussions and Philippe de Forcrand for helpful correspondence.

\end{document}